\documentclass[twocolumn,showpacs,preprintnumbers,amsmath,amssymb,superscriptaddress]{revtex4}

\usepackage{graphicx}
\usepackage{dcolumn}
\usepackage{bm}

\usepackage{amsfonts}
\usepackage{amssymb}
\usepackage{amsmath}
\usepackage{amsxtra}

\usepackage{epsfig}

\begin{document}

\title{Shell evolution and nuclear forces}

\author{N.~A.~Smirnova} 
\affiliation{CENBG (CNRS/IN2P3 - Universit\'e Bordeaux 1)
Chemin du Solarium, BP 120, 33175 Gradignan, France }
\author{B. Bally} 
\affiliation{CENBG (CNRS/IN2P3 - Universit\'e Bordeaux 1)
Chemin du Solarium, BP 120, 33175 Gradignan, France }
\author{K.~Heyde}
\affiliation{Department of Physics and Astronomy, University of Ghent,
Proeftuinstraat 86, B-9000 Ghent, Belgium}
\author{F.~Nowacki}  
\affiliation{Universit\'e Strasbourg, IN2P3, CNRS, F-67037 Strasbourg 2, France}
\author{K.~Sieja }
\affiliation{Institut f\"ur Kernphysik, Technische Universit\"at Darmstadt,
64289 Darmstadt, Germany and GSI-Hemholtzzentrum f\"ur
Schwerionenforschung, Planckstrasse 1, 64220 Darmstadt, Germany }

\date{\today}

\bibliographystyle{prsty}

\begin{abstract}
We present a quantitative study 
of the role played by different components characterizing the nucleon-nucleon interaction
in the evolution of the nuclear shell structure. 
It is based on the spin-tensor decomposition of an effective two-body shell-model interaction and 
the subsequent study of effective single-particle energy variations in a series of isotopes or isotones. 
The technique allows to separate unambiguously contributions of the central, vector and tensor components of
the realistic effective interaction.
We show that while the global variation of the single-particle energies is due to the central component
of the effective interaction, the characteristic behavior of 
spin-orbit partners, noticed recently, is mainly due to its tensor part.
Based on the analysis of a well-fitted realistic interaction in $sdpf$ shell model space,
we analyze in detail the role played by the different terms in the formation and/or disappearance
of $N\!=\!16$, $N\!=\!20$ and $N\!=\!28$ shell gaps in neutron-rich nuclei.
\end{abstract}


\bibliographystyle{prsty}

\maketitle

The shell structure is a common feature of finite quantum systems.
Amongst them, atomic nuclei represent unique objects characterized by the appearance
of a specific shell structure.
In particular, the magic numbers which correspond to the shell closures,
will change depending on the $N/Z$ ratio, i.e. when we move from nuclei in the vicinity of the 
$\beta $-stability line towards the particle driplines.
This has attracted a lot of attention nowadays because an increasing number of nuclei far 
from stability have become accessible experimentally (e.g., \cite{SorlinPorquet} and references therein). 
The hope to reach even more exotic 
nuclei demands for an improved  modelization, i.e. in the context of nuclear astrophysics.
Since the underlying shell structure determines nuclear properties in a major way, 
changes of nuclear shell closures and the mechanisms responsible for that 
should be much better understood.

Recently, the role of different components of the nucleon-nucleon (NN) interaction in the evolution
of the shell structure has been actively discussed.
Based on the analysis of the origin of a shell closure at $N\!=\!16$, Otsuka et al~\cite{OtFu01} 
have suggested that a central spin-isospin-exchange term,
$f(r) (\vec{\sigma} \cdot \vec{ \sigma }) (\vec{\tau } \cdot \vec{\tau })$ 
of the NN interaction plays a decisive role in the shell formation.

However, from a systematic analysis of heavier nuclei,
another conjecture has been put forward, namely, the dominant role played by the tensor force~\cite{OtSu05}.
The evidence is based on the comparison of the position of
experimental  one-particle or one-hole states in nuclei adjacent to semi-magic configurations 
with the so-called effective single-particle energies (ESPE's).
Within the shell-model framework, the latter ESPE's are defined~\cite{OtHo01} as one-nucleon separation energies
for an occupied orbital (or extra binding gained by addition of a nucleon to an unoccupied orbital)
evaluated from a Hamiltonian containing nucleon single-particle
energies (the bare single-particle energies
with respect to a closed-shell core) plus the monopole part of 
the two-body residual interaction~\cite{BaFr64,PoZu81}, i.e.
\begin{equation}
\label{hamiltonian}
\hat H_{mon} = \sum_{j,\rho } \epsilon_j^{\rho } \hat n_j^{\rho } + 
\sum_{j,j',\rho ,\rho '} V_{jj'}^{\rho \rho '} 
\frac{\hat n_j^{\rho } (\hat n_{j'}^{\rho '} - \delta_{jj'}\delta_{\rho \rho '})}
{(1+\delta_{jj'}\delta_{\rho \rho '})} \, ,
%
\end{equation}
where $j$ denotes a set of single-particle quantum numbers $(nlj)$ and $\rho $ refers 
to a proton ($\pi $) or to a neutron ($\nu $), $\hat n_j^\rho $ are particle-number operators.
$V_{jj'}^{\rho \rho '}$ are centroids of the two-body interaction
defined as~\cite{BaFr64,PoZu81,ZuDu95}
\begin{equation}
\label{centroid}
V_{jj'}^{\rho \rho '} = \frac{\sum_J \langle j_{\rho }j '_{\rho ' } |V| j_{\rho } j'_{\rho '} \rangle_{JM} 
(2J+1)(1+(-1)^J \delta_{jj'} \delta_{\rho \rho '})}
{(2j_{\rho }+1)(2j'_{\rho '}+1-\delta_{jj'} \delta_{\rho \rho '})} \, ,
\end{equation}
where  the total angular momentum of a two-body state $J$ runs over all possible values.

The monopole Hamiltonian represents a spherical
mean field extracted from the interacting shell model.
Its spherical single-particle states, or ESPE's, provide an important ingredient for the formation
of shells and interplay between spherical configurations and deformation in nuclei.
Large shell gaps obtained from a monopole Hamiltonian are a prerequisite
to obtain certain magic numbers. A reduction of the spherical shell gaps may lead to formation of
a deformed ground state, if the correlation energy of a given excited configuration 
and a decrease in the monopole part are large enough to make such an intruder excitation energetically favorable.
  
For example, the ESPE of the $\nu 0f_{7/2}$ orbital at $Z\!=\!8$, $N\!=\!20$ is the difference between 
total energy obtained, using Eq.~(\ref{hamiltonian}),
for $^{28}$O in its ground state and $^{29}$O with an extra neutron in the $0f_{7/2}$ state assuming
normal filling of the orbitals (normal filling is used throughout this work). 
Considering a series of isotopes or isotones, it is clear that 
ESPE's will experience a shift provided by the monopole part of the
proton-neutron matrix elements, mainly. The bigger the overlap of the proton and neutron
radial wave functions and the higher the $j$-values of the orbitals considered will lead, 
in general, to more drastic changes. 
In the present study we take into account the mass dependence of the
two-body matrix elements of the effective interaction according to the rule:
$V(A) = (A_{core}/A)^{1/3} V(A_{core})$.

From the analysis of the experimental data and the ESPE's it has been noticed~\cite{OtSu05,Smi05} 
that systematically 
\begin{equation}
\label{tensor}
|V_{j_> j'_<}^{\pi \nu } | >|V_{j_> j'_>}^{\pi \nu } |, \quad  
|V_{j_< j'_>}^{\pi \nu } | >|V_{j_< j'_<}^{\pi \nu } | ,
\end{equation}
where $j_>= l+1/2$ and $j_<= l -1/2$ are proton orbitals and  $j'_>= l'+1/2$ and $j'_<= l' -1/2$
are neutron orbitals. Thus, an extra attraction is manifested between 
generalized spin-orbit partners (proton $j = l+ 1/2$ and neutron $j' = l' - 1/2$ with $l \ne l'$ or vice versa).

This remarkable property is in line with the analytic relation valid for a pure tensor force~\cite{OtSu05},
i.e. using the above notation, $(2 j_> + 1)V^{\pi \nu }_{j_>j'} + (2 j_< + 1)V^{\pi \nu }_{j_<j'}=0 $.
To strengthen this idea, Otsuka et al.~\cite{OtSu05} have compared changes of the ESPE's in Ca, Ni and Sb isotopes,
as due to the tensor force only and estimating its strength as resulting  
from a ($\pi + \rho $)-exchange potential with a cut-off at 0.7 fm, with available experimental data.

This work has stimulated a large number of investigations using mean-field approaches 
~\cite{Dob06,OtMa06,BrownDug06,BrinkSt07,SuIk07,LeBe07,LongSag07,GrassoMa07,Colo07,LongSag08,ZouColo08,
ZalDob08,BartelBen08,TarLiang08,ZalSat09}.
It is worth noting that phenomenological interactions, such as Skyrme and Gogny force, 
most frequently used in mean-field calculations, did not include
a tensor term~\cite{BeHe03}.
Provided its importance, a tensor term should be introduced and the parameters re-adjusted,
what up to now, is not sa\-tis\-factorily reached yet
(see, e.g.~Ref.~\cite{SatWyss08,ZalSat09}).

However, importance of the tensor force within the shell model~\cite{OtSu05,Smi05} 
is mainly demonstrated in an empirical way.
It is evident, that the choice of the particular cut-off that was used to fix the strength of the tensor
force component plays a crucial role in obtaining quantitative result for shifts in the ESPE's as
presented in Fig.~4 of Ref.~\cite{OtSu05}.
It is also well known that the NN interaction is subjected to a strong renormalization before it can
be handled as an effective interaction in many-body calculations 
within a restricted model space~\cite{MHJ95}.
It is not straightforward to trace how the tensor component will become renormalized amongst the other terms 
contributing to the NN interaction.
Moreover, many shell-model interactions having high descriptive and predictive power were obtained
by a $\chi^2$-fit of two-body matrix elements to reproduce known experimental levels 
for a wide range of nuclei studied within a given model space (e.g. ~\cite{USD,GXPF1}). 
Even the effective interactions, maximally preserving their microscopic origin (based on a G-matrix), 
need further phenomenological correction (see  e.g., \cite{PoZu81,NoPo09}).
There is strong indication that inclusion of three-nucleon forces can  
heal the microscopically derived effective interaction, in particular, improve
its monopole part (see Ref.~\cite{Otsuka3N} and references therein for ab-initio studies). 
However, there are still 
no systematic calculations available up to date for many-nucleon systems either
within the shell model, or within the density-functional approach.
This is why the present study of the two-nucleon case is still of interest.

In spite of the indirect evidence at a two-body level~\cite{OtSu05}, 
up to now, the role played by the tensor force is not well determined.
For example, recent shell-model studies based on large-scale calculations 
using a realistic effective interaction in the heavy Sn nuclei region~\cite{Now07}  
conclude on the absence of a characteristic effect expected to result from a tensor force component.

In this Letter we present a quantitative study of the role
played by different components of the {\it effective interaction}. 
It is based on the spin-tensor decomposition of the two-body interaction,
which involves tensors of rank 0, 1 and 2 in spin and configuration space.
The procedure allows to separate the central, vector and tensor parts of the effective interaction.
The monopole properties of each component can be studied separately, elucidating unambiguously
its role in the shell evolution.
The method has already been applied in a similar context~\cite{UmMu04,UmMu06},
however, the authors used different effective interactions in smaller model spaces, 
concluding on a second-order tensor effect only. 
Contrary to these results, we put into evidence
an important first-order tensor effect in the present study.

A spin-tensor decomposition of the two-particle interaction has been known 
for many years~\cite{ElJa68,Kir73,SchTr76,KlKn77,Yoro80,BrRi85,OsSt92}.
In a given model space, a complete set of two-body matrix elements determines
the properties of nuclei ranging within this space.
For spin $1/2$ fermions (nucleons),
one can construct from their spin operators a complete set of linear operators
in a two-particle spin space:
$$
\begin{array}{c}
S^{(0)} =1, \, S^{(0)}_2 = \left[ \sigma_1 \times \sigma_2 \right]^{(0)}, \, 
S^{(1)}_3 = \sigma_1 + \sigma_2 \, , \\
S^{(2)}_4 = \left[ \sigma_1 \times \sigma_2 \right]^{(2)}, \,
S^{(1)}_5 = \left[ \sigma_1 \times \sigma_2 \right]^{(1)}, \,
S^{(1)}_6 = \sigma_1 - \sigma_2 \, ,
\end{array}
$$
By coupling the spin tensor operators with the corresponding rank tensors
in the configuration space one can construct scalar interaction terms. The most general two-body interaction 
can then be written as
%
%
\begin{equation}
\label{decomp2}
V(1,2) \equiv V = \sum_{k = 0,1,2} \left(S^{(k)} \cdot Q^{(k)}\right) = \sum_{k = 0,1,2} V^{(k)} . 
\end{equation}

Here, $V^{(0)}$ and $V^{(2)}$ represent the central and tensor parts of the effective NN interaction.
The $V^{(k=1)}$ term contains the so-called symmetric ($S^{(1)}_{i=3}$) and 
antisymmetric ($S^{(1)}_{i=5,6}$) spin-orbit operators~\cite{KlKn77}, which we will denote as LS and ALS, respectively. 
To obtain the matrix elements for the different multipole components in $jj$ coupling,
first, one transforms two-body matrix elements between normalized and antisymmetrized states
from $jj$ coupling to $LS$ coupling in the standard way.
The $LS$-coupled matrix elements of $V^{(k)}$ can be calculated from the $LS$ coupled matrix elements
of $V$ as
\begin{widetext}
\begin{eqnarray}
\langle (ab): L S, J M T M_T | V^{(k)} | (cd) : L' S', J M T M_T \rangle & = &
(2k + 1) (-1)^J 
\left\{ \begin{array}{ccc} L & S & J \\ S' & L' & k 
  \end{array} \right\}
\sum_{J'} (-1)^{J'} (2J'+1)
\left\{ \begin{array}{ccc} L & S & J' \\ S' & L' & k 
  \end{array} \right \}  \nonumber \\
& \times & \langle (ab): L S, J' M T M_T | V | (cd) : L' S', J' M T M_T \rangle  ,
\label{decomp}
\end{eqnarray}
\end{widetext}
where $a \equiv (n_a,l_a)$.
Finally, starting from the $LS$ coupled matrix elements of  $ V^{(k)}$, for each $k$, 
we arrive at a set of $jj$ coupled matrix elements to be used for further investigation. 
It is important  to note that for a given set of quantum numbers of two-body states,
the matrix elements of $V$ are a sum of the matrix elements of its three components $V^{(k)}$.

In addition, using projection operators, one can select different components of 
the effective interaction that connect two-nucleon states with specific values of
the total spin $S$, isospin $T$  and parity $(-1)^L$. 
Thus, based on the selection rules in $LS$-coupling, we can separate triplet-even (TE),
triplet-odd (TO), singlet-even (SE) and singlet-odd (SO) channels of the central part, 
as well as the even and odd channels of the symmetric spin-orbit and tensor part.

Next, we propose to study the evolution of the ESPE's in series of isotopes and/or isotones
induced by a given effective interaction and of its different multipole components.
The decomposition described above is applicable only when the model space contains 
all spin-orbit partners within a given oscillator shell. 
This limits the region of applicability to the lighter nuclei. However, many interesting 
observations can still be extracted.

In the present paper, we explore the effective interaction in $1s0d1p0f$ shell-model space
that reproduces very well the properties of stable as well as nuclei
further away from stability~\cite{NoPo09}.
\begin{figure}
  \includegraphics[height=.162\textheight]{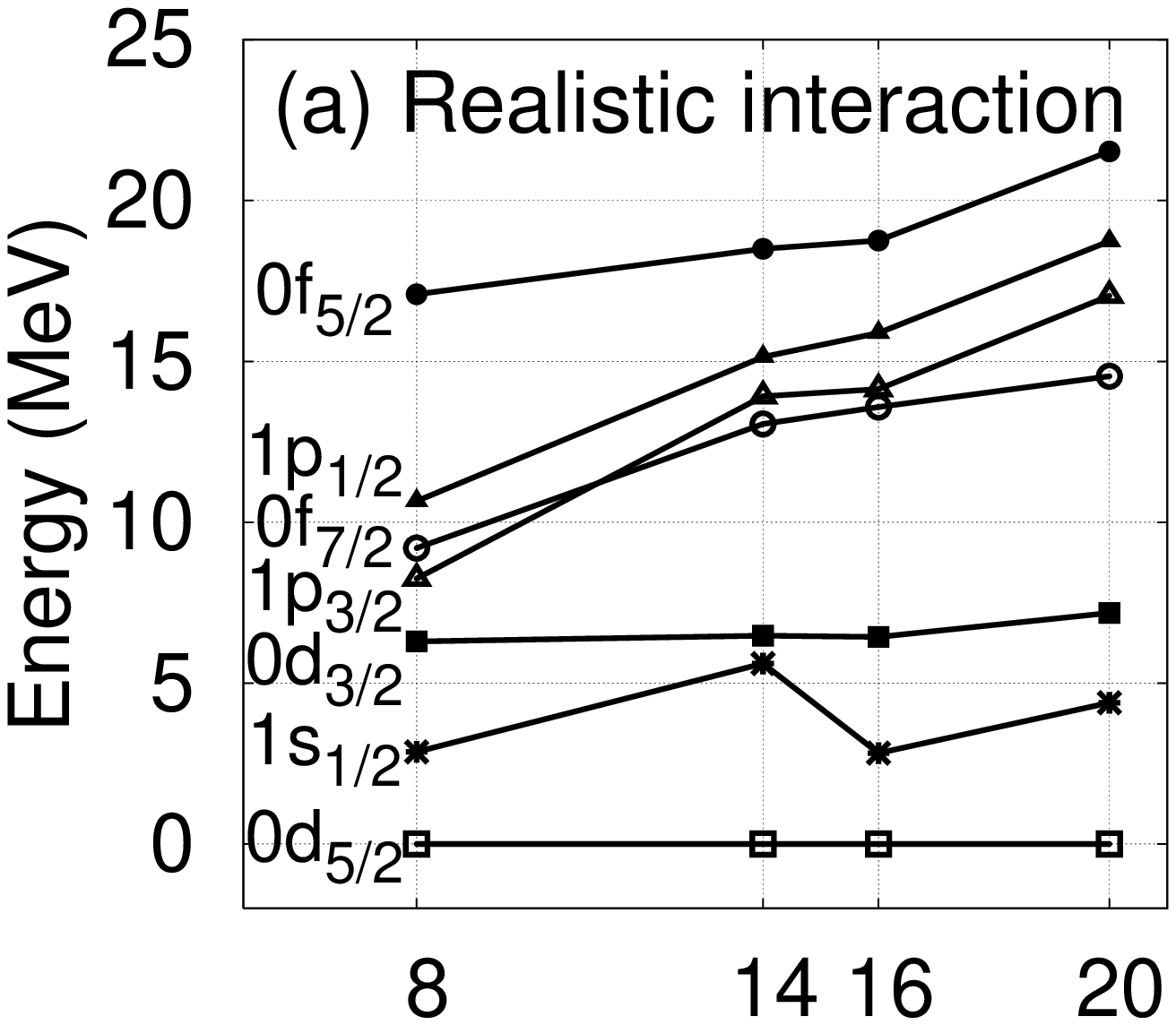} \hspace{-3mm}
  \includegraphics[height=.162\textheight]{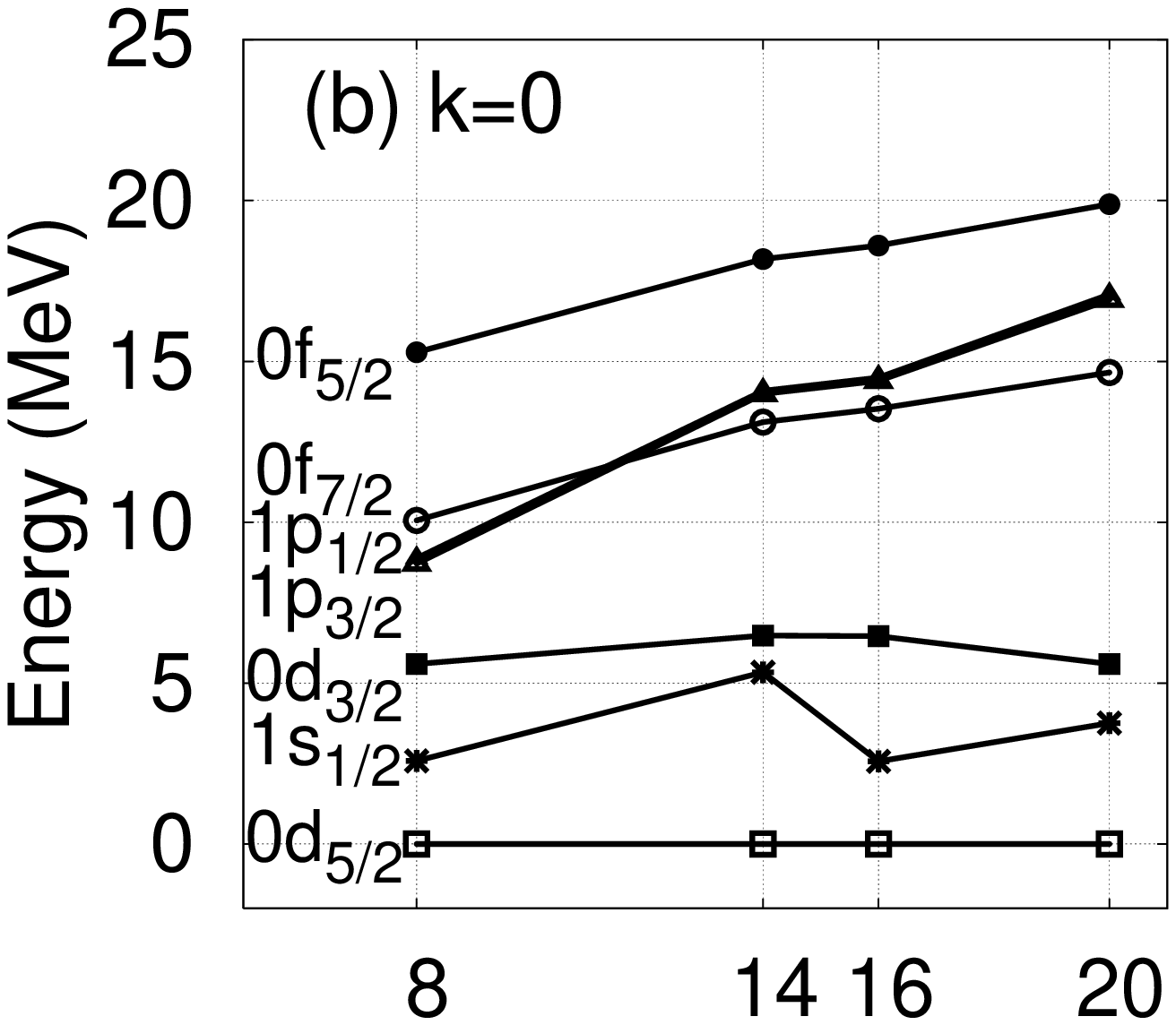} \\ 
  \includegraphics[height=.162\textheight]{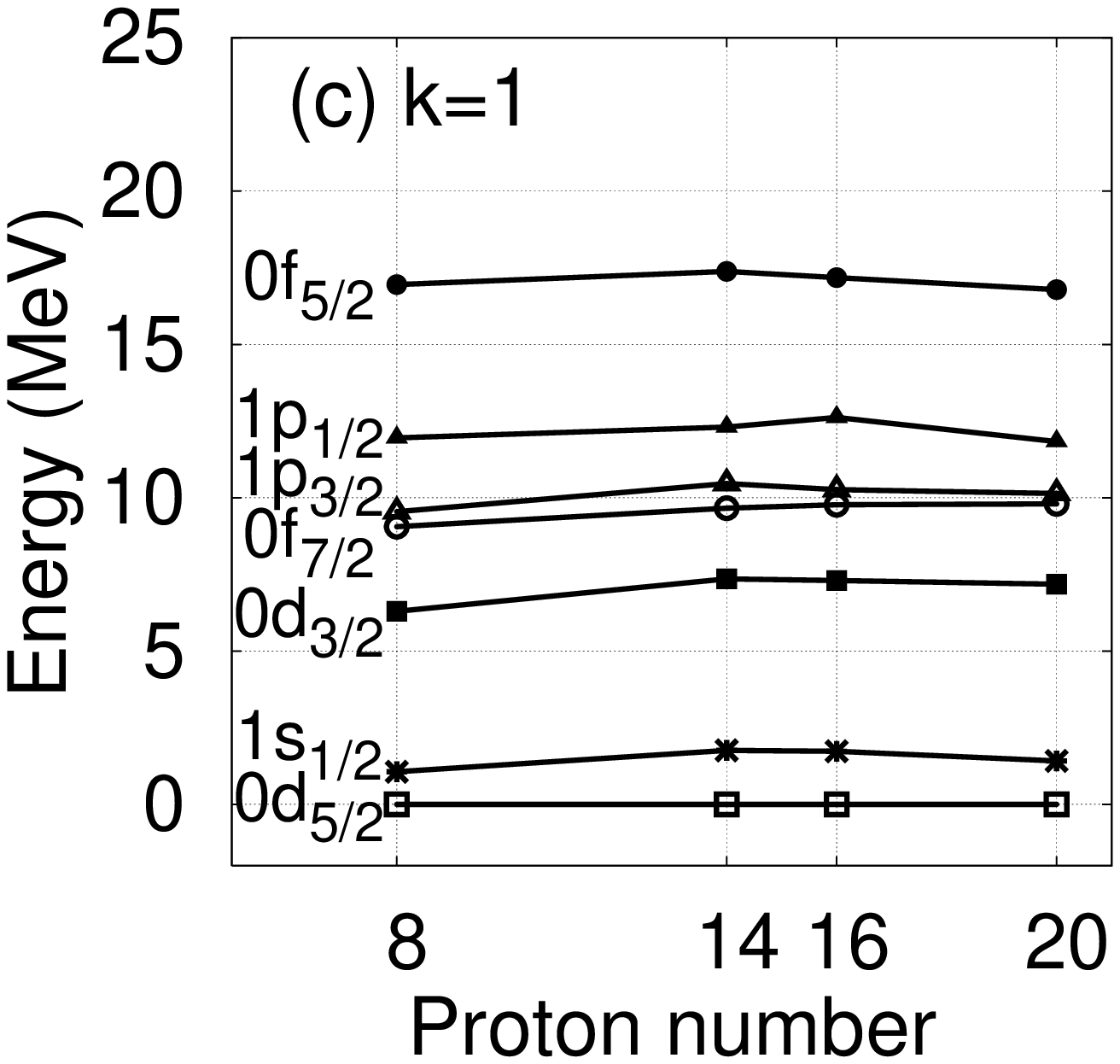} \hspace{-3mm}
  \includegraphics[height=.162\textheight]{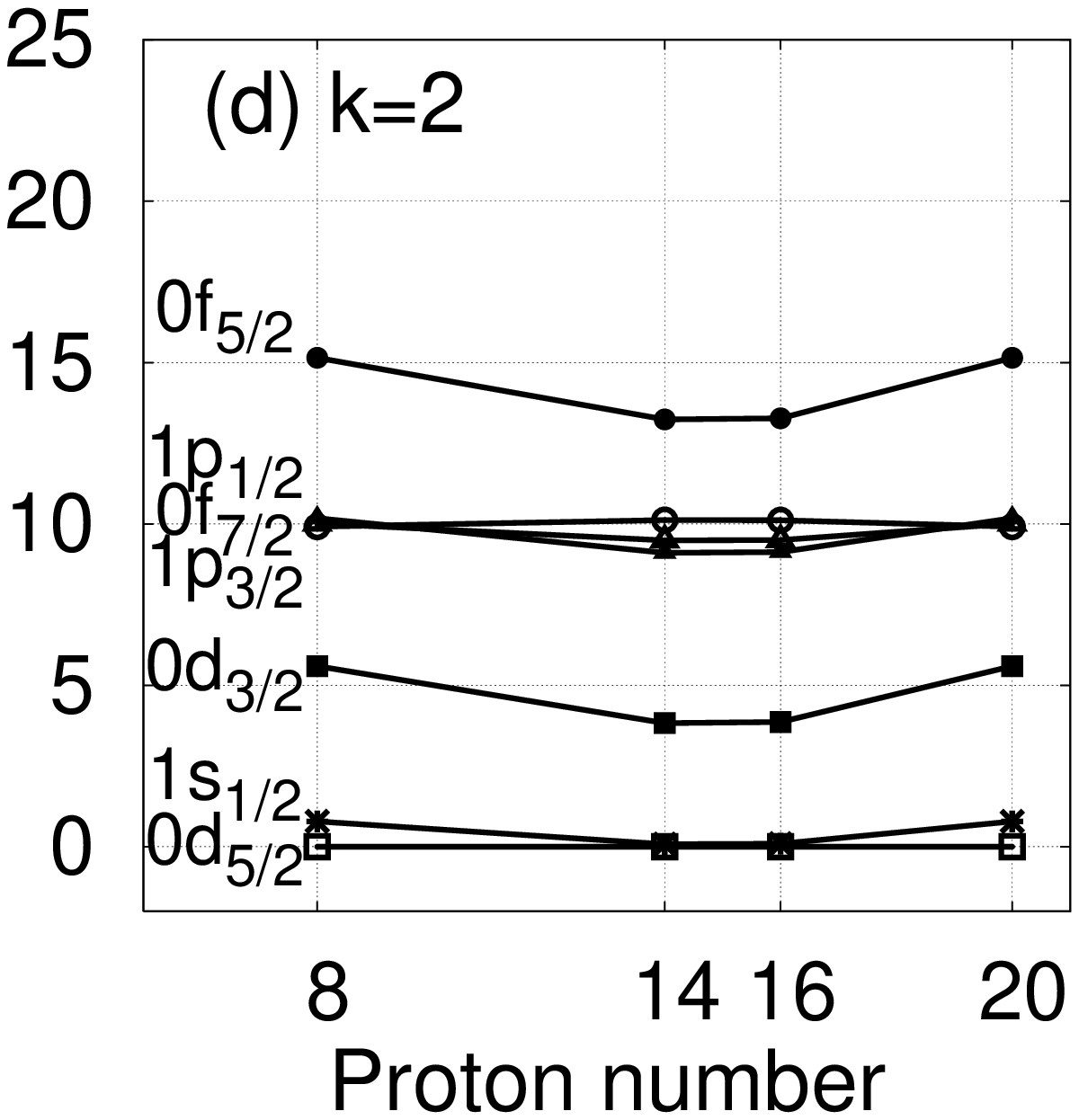} \\
  \caption{\label{fig:F-isotopes} Variation of the neutron ESPE's in $N\!=\!20$ isotones from O to Ca calculated 
using the realistic interaction~\cite{NoPo09} and its components separately.}
\end{figure}
We start with the ``classical'' case of the $N\!=\!20$ isotones, going  from O to Ca.
In Figure 1(a), we show the evolution of the neutron ESPE's using the realistic interaction~\cite{NoPo09}, 
relative to the energy of the $0d_{5/2}$ orbital.
When protons fill the $0d_{5/2}$ orbital (from O to Si),
the splitting between neutron $0f_{7/2}$ and $0f_{5/2}$ orbitals decreases. 
The opposite effect is observed when protons
fill the $0d_{3/2}$ orbital (from S to Ca) i.e. the corresponding splitting increases.
A similar but less pronounced behavior is noticed for the neutron $1p_{3/2}$ and $1p_{1/2}$ orbitals.
In Figures 1(b-d), we show the variation of the same ESPE's, 
this time caused by the $k=0$, 1 and 2 multipole components of the full interaction.
We remark that the summed shift of the energy for the various orbitals, produced by each two-body component, 
equals the total shift as produced by the two-body part of a full effective interaction, 
thus demonstrating the additivity.
We included in each figure the bare single-particle energies for visibility.

From Figure~1 one concludes that the central part of the effective interaction barely produces any relative
displacement of spin-orbit partners (Fig. 1(b)). A not very pronounced and often opposite effect
is induced by the $k=1$ part (Fig. 1(c)). It is indeed the tensor part which is responsible for the 
variation of spin-orbit partners, in line with the observation made in Ref.~\cite{OtSu05}.
\begin{table*}
\begin{widetext}
\centering
\begin{center} \caption{Contribution of different spin-tensor operators
to the energy splitting variations  $\Delta(j,j') \equiv \epsilon_j - \epsilon_{j'}$
in different regions: $N\!=\!20$ (columns 2-4), Ca-isotopes (columns 5-6), $N\!=\!28$ isotones (columns 7-8).} \vspace{2mm}
\begin{tabular}{cccccccc} 
\hline
Energy & $(\nu 0d_{3/2},\nu 1s_{1/2})$ &  $(\nu 0f_{7/2},\nu 0d_{3/2})$ &  
$(\nu 0f_{7/2},\nu 0d_{3/2})$ & $(\pi 0d_{3/2},\pi 0d_{5/2})$ & 
$(\pi 0d_{3/2},\pi 1s_{1/2})$ & $(\nu 1p_{3/2}, \nu 0f_{7/2})$ &  $(\nu 1p_{3/2},\nu 0f_{7/2})$ \\
gap & MeV & MeV & MeV & MeV & MeV & MeV & MeV \\
\hline
Filling  & $\pi 0d_{5/2}$ & $\pi 0d_{5/2}$ &  $\pi 0d_{3/2}$ & $\nu 0f_{7/2}$ & 
$\nu 0f_{7/2}$ &  $\pi 0d_{5/2}$ &  $\pi 0d_{3/2}$ \\
orbital  & $^{28}$O$\to ^{34}$Si & $^{28}$O$\to ^{34}$Si & $^{36}$S$\to ^{40}$Ca  &
$^{40}$Ca$\to $ $^{48}$Ca & $^{40}$Ca$\to $ $^{48}$Ca & $^{36}$O$\to ^{42}$Si & $^{44}$S$\to ^{48}$Ca  \\
\hline
Total      & {\bf -2.57} & {\bf 3.68} & {\bf 0.21} & {\bf -2.33} & {\bf -3.156} & {\bf 1.60} & {\bf 1.81}\\
\hline
Central    & {\bf -1.87} & {\bf 2.17} & {\bf 1.99} & {\bf -0.21} & {\bf -1.58} & {\bf 2.03} & {\bf 1.31} \\ 
	TE & -1.58 &  2.23 &  2.48 &  0.62 & -1.19 &  2.03 &  1.02 \\
	TO & -0.68 & -0.31 & -0.11 & -0.03 &  0.25 & -0.25 & -0.14 \\
	SE &  0.71 & -0.45 &  0.01 & -0.50 & -0.57 & -0.02 &  0.18\\
	SO & -0.32 &  0.70 & -0.39 & -0.30 & -0.07 &  0.28 &  0.25\\
\hline
Vector       &  {\bf 0.36} & {\bf -0.45} & {\bf  0.15} & {\bf 0.61} & {\bf 0.06} & {\bf 0.23} & {\bf -0.18} \\
	LS   & -0.05 & -0.10 & -0.16 &  0.09 & -0.15 &  0.11 &  0.15 \\
	even & -0.12 & -0.06 &  0.25 &  0.60 &  0.25 &  0.22 & -0.27 \\
	odd  &  0.07 & -0.04 & -0.41 & -0.51 & -0.40 & -0.11 &  0.41\\
	ALS  &  0.41 & -0.35 &  0.31 &  0.52 &  0.21 &  0.12 & -0.33 \\
\hline
Tensor      & {\bf -1.06} &  {\bf 1.96} & {\bf -1.93} & {\bf -2.73} & {\bf -1.64} & {\bf -0.67} &  {\bf 0.68}\\
	even & -0.78 &  1.31 & -1.28 & -1.59 & -0.96 & -0.43 &  0.43\\
	odd  & -0.28 &  0.66 & -0.65 & -1.14 & -0.68 & -0.24 &  0.26\\
\hline
\end{tabular}
\end{center}
\end{widetext}
\end{table*}
In Table~I we summarize contributions of the central, vector and
tensor terms in the spatial even and odd channels, separately. This elucidates the origin of 
the evolution of a given single-particle  energy splitting. The decrease of the splitting between  
$\epsilon (\nu 0d_{3/2})$ and $\epsilon (\nu 1s_{1/2})$ by 2.57 MeV, going from $^{28}$O to $^{34}$Si 
(Table~I, column 2), 
turns out to result from the combined effect of the central part (1.87 MeV), in particular, 
in its triplet-even channel,
and the tensor part of the nuclear interaction (1.06 MeV).

Similarly, the increase of the gap between the neutron $0d_{3/2}$ and $0f_{7/2}$ orbitals when
going from $^{28}$O to $^{34}$Si and onwards from $^{36}$S to $^{40}$Ca (columns 3 and 4
of Table I) is a joint effect of the central and tensor component of the effective interaction.
This is an important manifestation of the tensor force in this region. 
Due to the fact that at $N\!=\!20$ the above two neutron orbitals have (i) the same
radial quantum number, and, (ii) a different spin-to-orbital orientation, a large and negative tensor contribution 
of $-1.93$ MeV results for the variation of the gap between the $0d_{3/2}$ and $0f_{7/2}$ orbitals 
when filling the $0d_{3/2}$ orbital with protons (from  $^{36}$S to $^{40}$Ca). 
This large tensor shift, however, is almost fully cancelled by the central contribution of 1.99 MeV. 
The combined effect results in only a slight overall decrease of the $N\!=\!20$ shell gap from $^{40}$Ca to 
$^{36}$S and $^{34}$Si, thereby preserving the semi-magic nature of the latter nuclei.
At the same time, while filling the $0d_{5/2}$ orbital with protons (from $^{28}$O to $^{34}$Si), 
due to the change in the spin-to-orbital orientation with respect the proton $0d_{3/2}$ orbital,
the tensor contribution remains large but changes its sign (1.96 MeV). 
This enforces the central contribution (2.17 MeV) and results in a rapid decrease of the $N\!=\!20$ shell gap 
below $^{34}$Si which is at the origin of the so-called 'island of inversion' around $^{32}$Mg (deformed ground state).

The position of the $0d_{3/2}$ orbital and the possible shell gaps between this orbital and either the $1s_{1/2}$, 
or $0f_{7/2}$ orbital, plays an important role in the formation of $N\!=\!16$ as a magic number at $Z\!=\!8$. 
This was formerly ascribed to result from the spin-isospin exchange part 
of the central force component~\cite{OtFu01}
and sometime later to be due to mainly a pure tensor force~\cite{OtAb07}.
The present results support the important role of both a central part (in its spin-isospin exchange
channel) and a tensor part in changing the shell structure between O and Si.
\begin{figure}
  \includegraphics[height=.2\textheight]{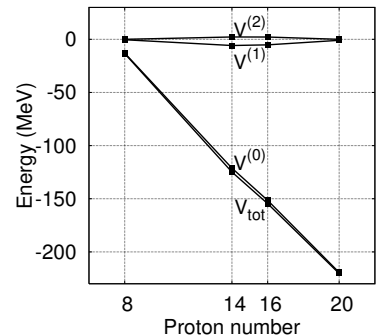}
   \caption{\label{fig:F-isotopes_d32} Two-body contribution to the binding energy of $^{27}$O, $^{33}$Si, 
   $^{35}$S and $^{39}$Ca in the lowest state (one hole
   in the $0d_{3/2}$ single-particle orbital), relative to the binding energy of  $^{16}$O ($N\!=\!20$ isotones) 
    using the realistic interaction~\protect\cite{NoPo09} and its components, separately.}
\end{figure}

In Figure 2, we show the two-body contribution to the binding energy from
the monopole part of the realistic interaction and its different components. 
For the analysis we choose the $N\!=\!19$ isotones $^{27}$O, $^{33}$Si, $^{35}$S and $^{39}$Ca with a neutron hole 
(relative to $N\!=\!20$) in the $0d_{3/2}$ state. As is seen, the global shift 
is due to the central part of the effective interaction when adding up to 12 protons to the $^{16}$O core. 
This contrasts with the results, presented before, in which we studied the local relative variations in 
the single-particle energy in which the tensor force component plays a major role.

\begin{figure}
 \includegraphics[height=.2\textheight]{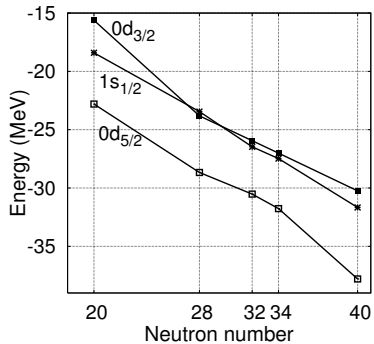}
  \caption{ Variation of proton single-hole states in
 Ca-isotopes using the realistic interaction \protect{\cite{NoPo09}}.}
\end{figure}

More evidence results from the single-proton holes in Ca-isotopes i.e. studying the K isotopes.
There is a crossing of the $1s_{1/2}$ and  the $0d_{3/2}$ orbitals when approaching $^{48}$Ca and,
in addition, a lowering of the energy gap between the $0d_{3/2}$ and the $0d_{5/2}$ orbitals when going from $^{40}$Ca to $^{48}$Ca,
confirmed experimentally~\cite{Doll76,Cottle98,SorlinPorquet}. 
In Figure~3, we show the variation of the proton ESPE's in Ca-isotopes obtained from the same effective interaction,
while in Table I we present a detailed analysis of the role of different components in
the evolution of the gaps.
It is seen (columns 5) that the lowering of the gap 
between proton $0d_{5/2}$ and $0d_{3/2}$ orbitals 
as neutrons fill the $0f_{7/2}$ orbital is mainly due to the tensor force.
However, it is the central part, combined with the contribution from the tensor force,
which reduces the gap between proton $0d_{3/2}$ and $1s_{1/2}$ orbitals when approaching $^{48}$Ca
(column 6). 

\begin{figure}
  \includegraphics[height=.2 \textheight]{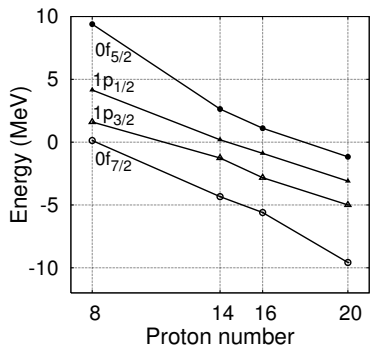}
  \caption{\label{fig:K-isotopes} Variation of neutron single-particle states in
 $N\!=\!28$ isotones using the realistic interaction \protect{\cite{NoPo09}}.}
\end{figure}

Finally, we explore evolution of neutron ESPE's in $N\!=\!28$ isotones, from O to Ca,
as a function of proton number. The characteristic trends of spin-orbit partners,
generic for a pure tensor force, are well manifested when using the same realistic interaction
(see Fig.~4):
approaching neutron $0f_{7/2}$ and $0f_{5/2}$ spin-orbit partners, and 
likewise for the  $1p_{3/2}$ and $1p_{1/2}$
spin-orbit partners, when filling the proton $0d_{5/2}$ orbital. 
An opposite effect results when filling  the $0d_{3/2}$ orbital and
fingerprints the contribution from a tensor term.

In Table I (columns 7 and 8), we analyze in detail the reduction of the $N\!=\!28$ shell gap, 
i.e. the change in the neutron  $1p_{3/2}$--$0f_{7/2}$ energy difference from $^{48}$Ca to the lighter isotones. 
To start with, these two orbitals have different radial quantum numbers. 
Therefore, the radial overlap contributing to the 
$V^{\pi \nu}_{ 0d_{3/2} 0f_{7/2}}$ centroid is larger than the radial overlap contributing to the
$V^{\pi \nu}_{ 0d_{3/2} 1p_{3/2}}$ centroid.
As can be seen from the table, the contributions from the central and tensor
terms are dominating. 
Since both the $0f_{7/2}$ and $1p_{3/2}$ orbital are 'spin-up' oriented ($j_>= l+1/2$),
the tensor term contributes in a similar way to the energy shift when protons fill the 
$0d_{3/2}$ orbital ($V^{(k=2) \pi \nu}_{ 0d_{3/2} 0f_{7/2}}$ 
and $V^{(k=2)\pi \nu}_{ 0d_{3/2} 1p_{3/2}}$ are both positive). 
The same happens when protons fill the $0d_{5/2}$ 
orbital ($V^{(k=2) \pi \nu}_{ 0d_{5/2} 0f_{7/2}}$ 
and $V^{(k=2)\pi \nu}_{ 0d_{5/2} 1p_{3/2}}$ are both negative).
The overall difference in sign is due to the different relative spin to orbital orientation of the 
neutron orbitals (both ($j{_>}^{\nu}= l^{\nu}+1/2$) relative to the proton orbitals ($j^{\pi}= l^{\pi} \pm 1/2$)).
Due to the difference in absolute value of the centroids, in particular, 
due to different radial overlaps  for $\nu 0f$-$\pi 0d$ versus $\nu 1p$ -$\pi 0d$,
the positive and negative tensor contributions to the energy centroid do not cancel. 
They result in a shift of 0.68 MeV and $-0.67$ MeV filling the $0d_{3/2}$ or $0d_{5/2}$ orbital, respectively.
Adding this tensor energy shift to the central plus vector energy shift results in a reduction of the $N\!=\!28$
shell gap going from $^{48}$Ca to $^{44}$S and from $^{42}$Si to $^{36}$O.  
This situation contrasts the $N\!=\!20$ shell gap evolution discussed before. 

These examples illustrate that in the discussion of shell gap evolution, it is mandatory
to take into account what particular orbitals are considered. Both the central and tensor term represent
important ingredients, together with the magnitude of the radial overlaps involved.

To summarize, we have proposed a quantitative study of the shell structure evolution
in series of isotopes or isotones, based on a spin-tensor decomposition of the two-body matrix elements.
The method has allowed us to clarify 
the role played by the different terms of the effective interaction in the variation of the
single-particle energy of different orbitals.

Based on the analysis of the best realistic interaction in the $1s0d1p0f$ shell model space~\cite{NoPo09},
we show that the evolution of the $N\!=\!16$, $N\!=\!20$ and $N\!=\!28$ shell gaps is a combined effect of different
spin-tensor terms, of which the central term in its triplet-even channel and the tensor term are of
overwhelming importance.
This conclusion partially supports the results of Ref.~\cite{UmMu04,UmMu06} regarding the importance
of the triplet even channel but evidences the crucial role of the first-order tensor term as conjectured in 
Refs.~\cite{OtSu05,Smi05}. 
The tensor term plays a dominant role, with increasing role of the vector term
in the single-particle energy difference for spin-orbit partners.
However, from the examples discussed here, one cannot assign unambiguously a dominating
role to the tensor mechanism in cases when no explicit spin-orbit partners are considered. 
For example,
the increase in energy splitting between proton $0h_{11/2}$ and $0g_{7/2}$ when filling neutron
$0h_{11/2}$ in heavy Sb isotopes, or in the energy splitting between neutron 
$0h_{11/2}$ and $0g_{7/2}$ when filling proton $0g_{9/2}$ in $N\!=\!51$ isotones, discussed in Ref.~\cite{OtSu05},
may be a result of different parts of the effective interaction. To clarify the observed situation, 
a corresponding quantitative analysis should be performed.

The decomposition is a suitable tool only for the model spaces when all spin-orbit partners are present.
This does not allow, at the present moment, to analyze heavier nuclei and check the hypothesis 
of the tensor force action in heavy systems, until a realistic determination of 
the position of relevant spin-orbit partners is established. 
As can be seen from the present work, the vector term of the effective interaction typically counter-acts
the tensor term. The increasing role of the vector term could form a plausible scenario 
for a reduction of the tensor effect in heavy nuclei
and explain the results obtained from shell-model studies of Ref.~\cite{Now07}.
This requires more data on key heavy nuclei and the availability of
extremely large-scale shell-model calculations.

N.A.S. thanks M.~Hjorth-Jensen for attracting attention to the existence
of a spin-tensor decomposition. 
We are grateful to O.~Sorlin for his interest and useful discussions.
K.H. thanks the FWO-Vlaanderen for financial support. This research was 
performed in the framework of the BriX network (P6/23) funded by the 
'IUAP Programme --- Belgian State-BSP'.
K.S. has been supported by the state of Hesse within
the {\it Helmholtz International center for FAIR} (HIC for FAIR)
and by the DFG under grant no. SFB 634.

\bibliography{decomp_new4}

\end{document}